\documentclass[reprint,aps,amsmath,amssymb,twoside,prd,showkeys,superscriptaddress]{revtex4-1}
\usepackage{verbatim}
\usepackage{graphicx}
\usepackage[T1]{fontenc}
\usepackage{textcomp}
\usepackage{txfonts}
\usepackage{tensor}
\usepackage{cancel}
\usepackage{color}
\newcommand{\HCd}{\mathcal{H}}

\def\HCdt0{\tilde{\HCd}_{0}}
\newcommand{\afffias}{Frankfurt Institute for Advanced Studies (FIAS), Ruth-Moufang-Strasse~1, 60438 Frankfurt am Main, Germany}
\newcommand{\affjwg}{Goethe-Universit\"at, Max-von-Laue-Strasse~1, 60438~Frankfurt am Main, Germany}
\newcommand{\affbgu}{Physics Department, Ben-Gurion University of the Negev, Beer-Sheva 84105, Israel}
\newcommand{\affbahamas}{Bahamas Advanced Study Institute and Conferences, 4A Ocean Heights, Hill View Circle, Stella Maris, Long Island, The Bahamas}
\preprint{}
\begin{document}
\title{Inflation Compactification from Dynamical space time}
\author{David Benisty}
\email{benidav@post.bgu.ac.il}
\affiliation{\afffias}\affiliation{\affjwg}\affiliation{\affbgu}
\author{Eduardo I. Guendelman}
\email{guendel@bgu.ac.il}
\affiliation{\afffias}\affiliation{\affbgu}\affiliation{\affbahamas}
\date{\today}
\begin{abstract}
A mechanism of inflation from higher dimensions compactification is studied. An Early Universe capable of
providing exponential growth for some dimensions and exponential
contraction for others, giving therefore an explanation for the big size of
the observed four dimensional Universe as well as the required smallness
of the extra dimensions is obtained. The mechanism is formulated in the context of dynamical space time theory which produces a unified picture of dark energy, dark matter and can also provides a bounce for the volume of the universe. A negative vacuum energy puts an upper bound on the maximum volume and the bounce imposes a lower bound. So that in the early universe the volume oscillates, but in each oscillation the extra dimensions contract exponentially, and the ordinary dimension expand exponentially. The dynamical space time theory provides a natural way to exit from the inflation compactification epoch since the scalar field that drives the vacuum energy can smoothly climb into small positive values of vacuum energy, which is the end of the inflation compactification. A semi analytic solution for a step function potential is also studied, where all of these effects are shown. Especially the jump of the vacuum energy affect only on the derivative of dynamical space time vector field, and not the volume or it's derivatives, which match smoothly. 
\begin{description}
\item[Key Words]
Inflation Compactification, Dynamical space time, Unified Dark energy Dark matter
\end{description}
\end{abstract}
\pacs{Valid PACS appear here}
\maketitle
\section{introduction}
In many interesting models a vacuum with negative cosmological constant is predicted, as in super-strings, super-gravity \cite{Freund:1980xh} etc. Here we will present a model which uses extra dimensions and a primordial negative vacuum energy. For cosmology with higher dimensions and in the presence non canonical scalar field whose dynamics is governed by a dynamical space time vector field, which is used as a Lagrange multiplier of an energy momentum tensor of the scalar field \cite{Benisty:2018qed}. A non Lagrangian approach similar to this was developed by Gao and collaborators \cite{Gao:2009me}. 

An interesting feature of the model \cite{Benisty:2018qed} is that it allows for bouncing solutions. This effect in higher dimensions combined with a negative cosmological constant in the early universe, leads to the existence of a "inflationary phase" for some dimensions and a simultaneous "deflationary phase" for the remaining dimensions, since the volume of the space time remains constant or oscillating in the early universe. For an approximately constant volume some dimensions will grow exponentially, where the others will shrink exponentially. This effect is obtained without invoking exotic matter or quantum effects as a similar  inflation compactification scenario was discussed in \cite{Guendelman:1990kg}\cite{Guendelman:1993ty}\cite{Guendelman:2003tm}\cite{Ho:2010vv}\cite{Szydlowski:1990ph}.   

We discuss how it may be possible to exit from this inflation-compactification era by dynamically increasing the cosmological constant, until it becomes positive and small. This is possible in our model because the scalar field can evolve towards increasing values of vacuum energy, without problems.  Finally the need for trapping the extra dimensions when they become become very small to prevent their re-expansion is studied. One could use for example the Casimir effect for periodic extra dimensions for this purpose. 

\section{The Basis of the mechanism}
\label{sec:intro}  
\subsection{The geometry}
For understanding the basics of the  mechanism, we review the formalism of cosmology with higher dimension as developed in \cite{Tosa:1984gr} for a "Classical Kaluza-Klein cosmology for a torus space with a cosmological constant and matter". The metric we assume is the following: 
\begin{equation}\label{metric}
ds^2 = -dt^2 + R(t)^2 \frac{\Sigma dx^j dx^j}{f_D(x)^2} + R(t)^2 \frac{\Sigma dx^p dx^p}{f_d(x)^2}
\end{equation}
where:
\begin{equation}
f_D = 1 +\frac{k_D}{4}\Sigma (x^i)^2, \quad f_d = 1 +\frac{k_d}{4}\Sigma (y^p)^2
\end{equation}
$R(t)$ is the scale factor for the $D$ dimensions ($x^i$) and $r(t)$ is the scale factor for the other $d$ dimensions ($y^p$). The $k_d$ and $k_D$ are the special curvatures. Their Hubble constants are defined as:
\begin{equation}
\mathcal{H}_R = \frac{\dot{R}}{R} \quad \mathcal{H}_r = \frac{\dot{r}}{r}
\end{equation}
The complete volume of the universe is defined as:
\begin{equation}\label{volume}
V = R^D r^d
\end{equation}
which allows us to define the "volume expansion parameter":
\begin{equation}\label{HC}
\mathcal{H} = \frac{\dot{V}}{V} 
\end{equation}
The connection between the volume expansion parameter and the Hubble parameters, using the definition (\ref{volume}) is:
\begin{equation}
\label{HC}
\mathcal{H} = D \mathcal{H}_R + d \mathcal{H}_d 
\end{equation}
The motivation for defining the total volume is because of the ability to write down one combination of Einstein equation which has no dependence on the individual scale factor, only through the volume, as we will see below.  
\subsection{Einstein equations}
We first consider the case of a stress energy tensor which has for every individual scale factor has it's own pressure: $p$ for $D$ dimensions, and $p'$ for $d$ dimensions:
\begin{equation}\label{set}
T^{\mu}_{\nu}=\textbf{diag}(\rho,-p,-p,...,-p',-p',...)
\end{equation}
Using the identities from the Appendix, we can obtain the solution for the Einstein equation:
\begin{equation}\label{density1}
\begin{aligned}
\frac{1}{2}D(D-1)\left[\frac{\dot{R}^2}{R^2} + \frac{k_D}{R^2}\right] +\frac{1}{2}d(d-1)\left[\frac{\dot{R}^2}{R^2} + \frac{k_d}{R^2}\right] \\ + D d \frac{\dot{R}}{R}\frac{\dot{r}}{r} = 8\pi \rho
\end{aligned}
\end{equation}
\begin{equation}\label{pressure1}
\begin{aligned}
(D-1)\frac{\ddot{R}}{R} + d\frac{\ddot{r}}{r} - d \frac{\dot{R}}{R}\frac{\dot{r}}{r} - (D-1) \left[ \frac{\dot{R}^2}{R^2} + \frac{k_D}{R^2} \right] \\ = -8\pi (\rho + p)
\end{aligned}
\end{equation}
\begin{equation}\label{pressure2}
\begin{aligned}
D\frac{\ddot{R}}{R} + (d-1)\frac{\ddot{r}}{r} - D \frac{\dot{R}}{R}\frac{\dot{r}}{r} - (d-1) \left[ \frac{\dot{r}^2}{r^2} + \frac{k_d}{r^2} \right] \\ = -8\pi (\rho + p')
\end{aligned}
\end{equation}
For simplicity we set the special curvature for all the dimensions to zero $k_d=k_D=0$. Under the assumption of isotropy of the pressure $p = p'=(\gamma-1) \rho$, the relation from Eqs. (\ref{density1})-(\ref{pressure2}) gives:
\begin{equation}
D\frac{\ddot{R}}{R}+d \frac{\ddot{r}}{r} =\frac{8\pi\rho }{D+d-1} [1-(D+d)\gamma]
\end{equation}
By the definition of the volume (\ref{volume}) the equation could be represent as:
\begin{equation}
\frac{\ddot{V}}{V} = \frac{D+d}{D+d-1} 8\pi (\rho-p)
\end{equation}
The notation of normalized density gives a dimensionless equation of motion. By integrating the equation and using the dimensionless density $\Omega := \frac{\rho}{\rho_c}$, we obtain the as in \cite{Tosa:1984gr}: 
\begin{equation}\label{EE}
E=\frac{1}{2}\dot{V}^2-\frac{D+d}{D+d-1} \Omega V^2
\end{equation}
where $E$ is the anisotropy parameter which is an integration constant that appears in the solution. The special feature of this equation is that it depends of the total volume and not by the separate scale parameters of the individual dimensions.

Using the volume definition again (\ref{volume}) in the Energy equation  (\ref{EE}) together with (\ref{density})  we obtain the first-order differential equations for $R$
and $r$ in terms of the volume solution:
\begin{subequations}
\begin{equation}
\frac{\dot{R}}{R} = \frac{1}{(D+d)V}[\dot{V}+\sqrt{\frac{2Ed}{D}(D+d-1)}]
\end{equation}
\begin{equation}
\frac{\dot{r}}{r} = \frac{1}{(D+d)V}[\dot{V}-\sqrt{\frac{2ED}{d}(D+d-1)}]
\end{equation}
\end{subequations}
From those equations we obtain that the basic condition for existence of solution is $E \neq 0$, because of it's appearance of the square root of $E$. After an integration:
\begin{subequations}\label{sf}
\begin{equation}
R(t)=V^\frac{1}{D+d} \exp[+\frac{1}{D+d}\sqrt{\frac{2Ed(D+d-1)}{D}}\int{\frac{dt}{V}}]
\end{equation}
\begin{equation}
r(t)=V^\frac{1}{D+d} \exp[-\frac{1}{D+d}\sqrt{\frac{2ED(D+d-1)}{d}}\int{\frac{dt}{V}}]
\end{equation}
\end{subequations}
Those equations could be used for obtaining the solution for every individual scale parameter of any particular dimension. After calculating the solution for the total volume from the energy equation, the equations above could give us the evolution of each scale factor. Notice that for any $E > 0 $ we get an anisotropic evolution.
\begin{figure*}[t]
 	\centering
 \includegraphics[width=0.9\textwidth]{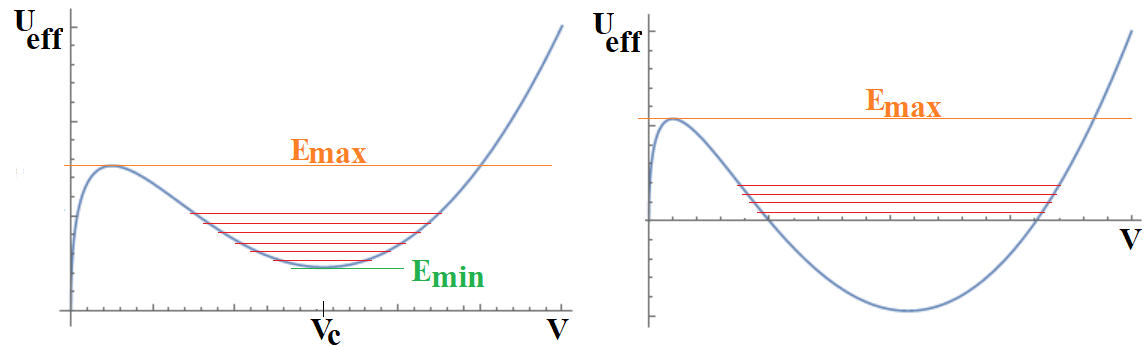}
\caption{The effective potential, for two cases, where $\Omega_\Lambda,\Omega_\kappa$<0 and $\Omega_m>0$.}
 	\label{fig1}
 \end{figure*} 
 \subsection{Solutions with constant equation of state}
Simple examples of density dependence on the volume could be given under the assumption of constant equation of state $\omega = \frac{p}{\rho}$. 
\begin{table}[h]
  \begin{center}
  \label{tab:table1}
    \begin{tabular}{l|c|r} 
    \textbf{Name} & $\omega$ & $\,\rho$ \textbf{dependence}\\
       \hline
      Stiff & $1$ & $V^{-2}$\\
      matter &  $0$  & $V^{-1}$\\
      radiation & $\frac{1}{D+d} $  & $V^{-\frac{D+d+1}{D+d}}$ \\
      dark energy & $-1$ & Constant\\
    \end{tabular}
    \caption{The properties of densities as a function of the volume}
  \end{center}
\end{table}
Substituting the density which the universe contains into the energy equation, would give the density multiple by the quadratic term of the volume. That means, for massless scalar field $\rho_\phi V^2$ will provide a constant term into the energy equation. However only for a ghost kinetic term of the scalar will shift the potential upwards, and for a physical kinetic term of a scalar will push the potential down.
The contribution for dark energy will be a parabolic term $V^2$ into the effective potential of the energy equation. If the dark energy has negative values the parabola will provide a barrier, which will prevent high values of $V$. A positive value of dark energy will provide a non-stable effective potential, which pushes the universe to infinity. The dark matter gives a linear term in the effective potential of the energy equation with a negative slope.

\subsection{Kasner solution}
For a complete vacuum free from density and pressure, the Kasner solution automatically follows from the basic formalism. From the energy equation (\ref{EE}) we get $V = \sqrt{2E} t$ which means the total volume grows linearly with the time. This case leads to well known Kasner vacuum solution \cite{Kasner} which describes an anisotropic universe without matter, with different scale factors as well. Using Eq (\ref{sf}) we get the powers for the scale factors $R(t) = t^p$, $r(t) = t^q$:
\begin{equation}
p = \frac{1}{D+d} (1+\sqrt{\frac{d(D+d+1)}{D}})
\end{equation}
\begin{equation}
q = \frac{1}{D+d} (1-\sqrt{\frac{D(D+d+1)}{d}})
\end{equation}
which obey the Kasner conditions:
\begin{equation}
D p + d q = Dp^2 +dq^2 =1
\end{equation}
This solution allows us to check that our formalism recovers the well known vacuum solutions.
\section{Inflation from Unified DE-DM}
\subsection{Unified Dark Energy and Dark Matter solution}
A suggestion for an action which produces DE-DM unification takes the form of \cite{Benisty:2018qed}:
\begin{equation}
\mathcal{L}=-\frac{1}{2}R+\chi_{\mu;\nu} T^{\mu\nu}_{(\phi)} - \frac{1}{2}g^{\alpha\beta} \phi_{,\alpha}\phi_{,\beta} - V(\phi)
\end{equation}
where $R$ is the Ricci scalar ($8 \pi G=1$), $\phi$ is a quintessential scalar field with a potential $V(\phi)$, and $\chi_\mu$ is a Dynamical space-time vector field which is a Lagrange multiplier enforces the covariant conservation law of the energy momentum tensor:
\begin{equation}
\nabla_\mu T^{\mu\nu}_{(\phi)} =0
\end{equation}
We use the same stress energy tensor as the one postulated by Gao and colleagues \cite{Gao:2009me}:
\begin{equation}\label{tmunuchi}
T^{\mu\nu}_{(\phi)} = -\frac{1}{2} \phi^{,\mu} \phi^{,\nu} + U(\phi) g^{\mu \nu}  
\end{equation}
which they require to be conserved without an action principle. The covariant conservation of this stress energy tensor lead to unified Dark Energy and Dark Matter for a constant potential, and for interacting DE-DM for non constant potential $U(\phi)$. This action produces very similar effects, but include additional effects like bouncing, which are not obtained in \cite{Gao:2009me}. The action depends on three different variables: the scalar field $\phi$, the dynamical space time vector $\chi_\mu$ and the metric $g_{\mu\nu}$.

\subsection{Equations of motion}
According to this ansatz, the scalar field is just a function of time $\phi(t)$ and the dynamical vector field will have only the time component $\chi_\mu = (\chi_0,0,0,0)$, where $\chi_0$ is also just a function of time.
A variation with respect to the dynamical space time vector field $\chi_\mu$ will force a conservation of the original stress energy tensor, which implies:
\begin{equation}\label{1frw}
\ddot{\phi}+\frac{1}{2}\mathcal{H}\dot{\phi}+U'(\phi)=0
\end{equation}
Notice that for a standard quintessence, the equation does not contain the factor $\frac{1}{2}$, but only the "volume expansion parameter" $\mathcal{H}$ which equals to $3H$ for the case of isotopic expansion of 3+1 dimensions. The second variation with respect to the scalar field $\phi$ gives a non-conserved current: 
\begin{subequations}\label{current}
\begin{equation}
\chi^\lambda_{;\lambda} U'(\phi) - V'(\phi) = \nabla_\mu j^{\mu} 
\end{equation}
\begin{equation}
j^{\mu}  = \frac{1}{2}\phi_{,\nu} (\chi^{\mu;\nu}+\chi^{\nu;\mu}) + \phi^{,\mu}
\end{equation}
\end{subequations}
and the derivatives of the potentials are the source of the current. For constant potentials the source becomes zero, and we get a covariant conservation of this current. For the metric we presented above, this equation of motion takes the form:
\begin{equation}\label{chi}
\ddot{\phi}(\dot{\chi}_0 - 1)+\dot{\phi}(\mathcal{H}(\dot{\chi}_0 - 1) + \ddot{\chi}_0) = U'(\phi)(\dot{\chi}_0+\mathcal{H}\chi_0) - V'(\phi)
\end{equation}
For constant potentials the current (\ref{current}) is covariantly conserved, a feature which will be used later. The last variation, with respect to the metric, gives the stress energy tensor we know from Einstein equation:
\begin{equation}\label{setv}
\begin{split}
G^{\mu\nu}= g^{\mu\nu} (\frac{1}{2}\phi_{,\alpha} \phi^{,\alpha} +  V(\phi)
+\frac{1}{2}\chi^{\alpha;\beta}\phi_{,\alpha} \phi_{,\beta}+\chi^{\lambda}\phi_{,\lambda}U'(\phi)) 
\\ 
- \frac{1}{2} \phi^{,\mu} ((\chi^{\lambda}_{;\lambda}+2) \phi^{,\nu} + \chi^{\lambda;\nu}\phi_{,\lambda} + \chi^\lambda \phi^{,\nu}_{;\lambda})
\\ - \frac{1}{2}(\chi^{\lambda}\phi^{,\mu}_{;\lambda}\phi^{,\nu}+\chi^{\lambda;\mu}\phi_{,\lambda}\phi^{,\nu})
\end{split}
\end{equation}

For the stress energy tensor from Eqs.(\ref{set}) and (\ref{setv}), the relation between the energy density and the fields is:
\begin{equation}\label{density}
\rho = (\dot{\chi}_0-\frac{1}{2})\dot{\phi}^2 + V(\phi)
\end{equation}
which has no dependence on the potential $U(\phi)$ or it's derivatives. Those three variation are sufficient for building a complete solution of the theory, using the energy equation (\ref{EE}) and the integration form of the individual scale parameters (\ref{sf}).

\subsection{Constant potentials solution}

In order to compute the evolution of the scalar field and we have to specify a form for the potentials. For a simplified case of constant potentials:
\begin{equation}
U(\phi)= C ,\quad  V(\phi)=\Omega_\Lambda
\end{equation}
The solution for the variation with respect the dynamical time, which is equation (\ref{1frw}) can be integrated to give:
\begin{equation}\label{eom1f}
\dot{\phi}^2 = \frac{2\Omega_m}{V}
\end{equation}
where $\Omega_m$ is an integration constant which represent the effective dark matter ratio. From the second variation, with respect to the scalar field $\phi$ a conserved current is obtained, which from equation (\ref{chi}) gives the exact solution of the dynamical time vector field:
\begin{equation}\label{chisolution}
\dot{\chi}_0 = 1 + \frac{\kappa}{V^2} 
\end{equation}
where $\kappa$ is another constant of integration. Together with Eq. (\ref{eom1f}) and Eq (\ref{chisolution}) into the density equation (\ref{density}), the  volume dependence of $\Omega:=\frac{\rho}{\rho_c}$ is:
\begin{equation}
\Omega = \Omega_{\Lambda}+ \frac{\Omega_m}{V} + \frac{\Omega_\kappa} {V^{3/2}}
\end{equation}
\begin{figure}[h]
 	\centering
 \includegraphics[width=0.5\textwidth]{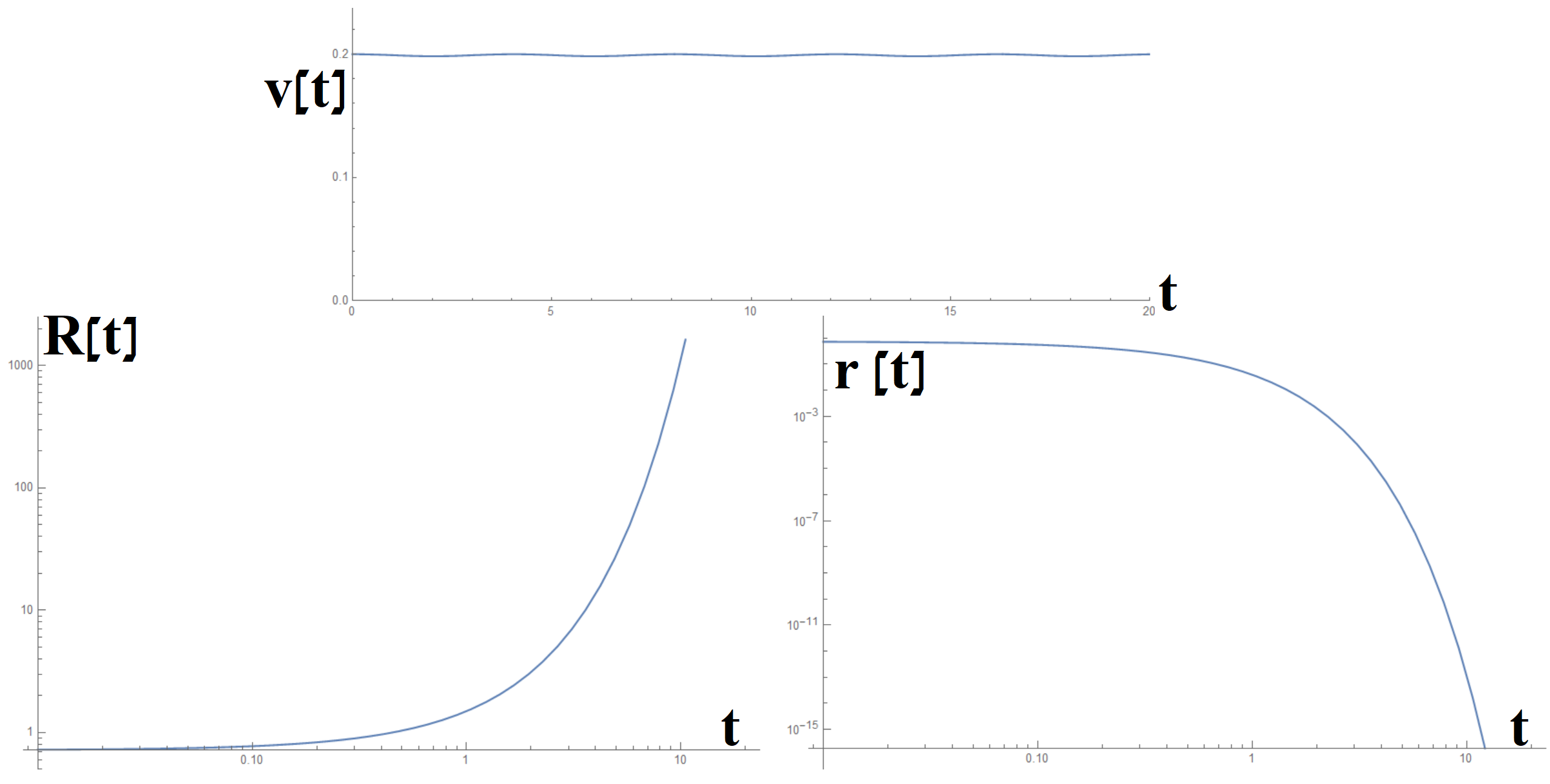}
\caption{A numerical solution for the volume and the scale factors in a Kaluza Klein universe, with the parameters: $\Omega_\Lambda = -0.04$, $\Omega_m = 0.24$, $\Omega_\kappa = -0.2$, with the initial condition $\dot{V} = 0.01$.}
 	\label{fig2}
 \end{figure}
where $\Omega_\kappa = \kappa \Omega_m$. Using the energy equation (\ref{EE}) which is the way to solve Einstein equations under the "inflation compactification mechanism", we obtain the relation: 
\begin{equation}\label{EE1}
E=\frac{1}{2}\dot{V}^2+U_{\textbf{eff}}(V)
\end{equation}
with the appropriate effective potential: 
\begin{equation}
U_{\textbf{eff}}(V) = -\frac{D+d}{D+d-1}(\Omega_{\Lambda} V^2 + \Omega_m V + \Omega_\kappa \sqrt[]{V})
\end{equation}
In figure (\ref{fig1}) we can see the plot of the effective potential for $\Omega_\Lambda,\Omega_\kappa<0$ and $\Omega_m>0$.

From equations ($\ref{sf}$) we can see terms with $\sqrt{E}$, therefore $E>0$ is a basic condition for existence of solutions, where $E$ is the measure of the anisotropy of the solution. Only for $E=0$ we have an isotropic solution. Because of this condition, we can obtain two different cases, represented in Figure (\ref{fig1}): the left case, where all the effective potential is positive every where, and the right case, where there is a part with negative values of the potential.

\begin{figure}[h]
 	\centering
 \includegraphics[width=0.5\textwidth]{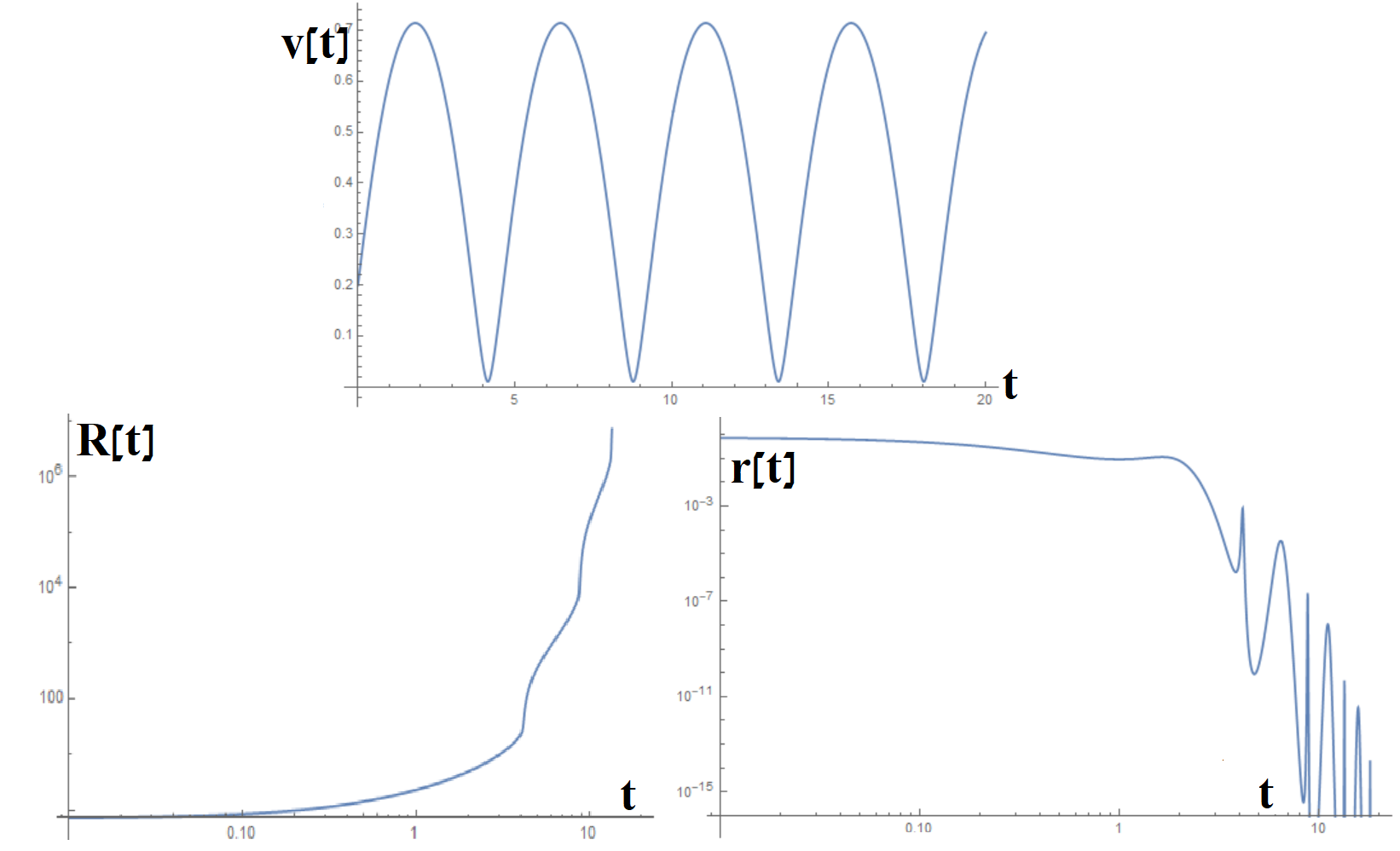}
\caption{A numerical solution of the volume and the scale factors for Kaluza Klein universe, with the parameters: $\Omega_\Lambda = -0.04$, $\Omega_m = 0.24$, $\Omega_\kappa = -0.2$ with initial condition $\dot{V} = 0.5$.}
 	\label{fig3}
 \end{figure}

In the left case, if $E=E_{min}$ we have $V=V_C=\textbf{const}$, which refers to a constant total volume. But from equation (\ref{sf}) we obtain that the scale parameter $R(t)$ is exponentially growing, and the $r(t)$ is exponentially shrinking:
\begin{subequations}\label{gs}
\begin{equation}
R(t)=V_C^\frac{1}{D+d} \exp[+\frac{1}{D+d}\sqrt{\frac{2Ed(D+d-1)}{D}}{\frac{t}{V_C}}]
\end{equation}
\begin{equation}
r(t)=V_C^\frac{1}{D+d} \exp[-\frac{1}{D+d}\sqrt{\frac{2ED(D+d-1)}{d}}\frac{t}{V_C}]
\end{equation}
\end{subequations}
This kind of solutions holds only for the left case, because of the energy condition $E=E_{min}>0$ could only exist if the potential is positive at the minimum. In general, when $E>E_{min}$ we have an oscillating volume solution. If $E$ is slightly larger than $E_{min}$ the oscillation won't be so large, and the expansion of the individual scale parameters will be close to an exponentially growing or decreasing, as shown in figure (\ref{fig2}). On the other hand, if $E$ is much larger than $E_{min}$, the oscillations will be large also, and the individual scale parameters will grow and shrink modulated by an oscillatory behavior, as shown in Figure (\ref{fig3}). Another important condition is $E<E_{max}$ as can be seen in Figure (\ref{fig1}). 

For proving the existence solutions for more non constant potential $V(\phi)$, where dynamically we change from negative to positive values, we study the case of step function potential.  
\subsection{End of inflation compactification, using a step function potential}
The end of the inflation compactification era will take place when the cosmological constant changes from negative values to positive values. Since then the effective potential does not prevent the total volume from expanding to infinity. For example a smooth potential that interpolates from those values is:
\begin{equation}
V(\phi) = \frac{\Lambda_{+\infty}-\Lambda_{-\infty}}{2} \tanh(\beta \phi) +\frac{\Lambda_{+\infty}+\Lambda_{-\infty}}{2}
\end{equation}
where $\Lambda_{+\infty}>0$ is the asymptotic value of the potential for $\phi \rightarrow \infty$, and is chosen to be small. On the other hand $\Lambda_{-\infty}<0$ is the asymptotic value of the potential for $\phi \rightarrow -\infty$.
For obtaining a partially analytic and more simple solution we take the limit for
$\beta \rightarrow \infty$, which then becomes a step function:
\begin{equation}\label{poten}
V(\phi) = \frac{\Lambda_{+\infty}-\Lambda_{-\infty}}{2} \textbf{Sign} (\phi) +\frac{\Lambda_{+\infty}+\Lambda_{-\infty}}{2}
\end{equation}
Notice that there is no problem for the scalar field $\phi$ to increase and go up in the direction of increasing dark energy, since it's dynamics is not determined by the potential $V(\phi)$, that determined the value of the dark energy. In general the scalar field evolution depend on $U'(\phi)$, which is in this case zero, since the potential $U(\phi)$ is a constant. Still by choosing the positive root of Eq. (\ref{eom1f}) we get the desired effect of increasing dark energy as a function of time.
For simplicity let's define the parameter $\xi$:
\begin{equation}
\xi=\dot{\chi}_0-1
\end{equation}
which estimates the difference between the dynamical time and the cosmic time. If $\xi=0$ then $\chi_0 = t$. The variation with respect to the scalar field $\phi$, Eq. (\ref{chi}), takes the form: 
\begin{equation}
\dot{\phi}(\dot{\xi}+\frac{1}{2}\mathcal{H}\xi)=-V'(\phi)
\end{equation}
Since $\phi$ is a monotonic function of time, it's better to change the time dependence to the scalar dependence $\frac{d}{dt} = \dot{\phi} \frac{d}{d\phi}$. In this way, the equation is easier to analyze:
\begin{equation}
\frac{2\Omega_m}{V^{\frac{3}{2}}} d(\xi V^{\frac{1}{2}}) = -d V(\phi)
\end{equation}
For the potential (\ref{poten}) we get the differential equation: 
\begin{equation}
\frac{2\Omega_m}{V} (\frac{d}{d\phi}\xi+\frac{1}{2V} \frac{d}{d\phi} V)=-(\Lambda_{+\infty}-\Lambda_{-\infty}) \delta(\phi)
\end{equation}
where the in the right hand side we obtain a source term, with the piecewise  solution:
\begin{equation}\label{xil}
\xi(\phi <0) = \frac{\kappa_{-}}{V_{(\phi=0)}^{\frac{1}{2}}} \quad , \quad \xi(\phi >0) = \frac{\kappa_{+}}{V_{(\phi=0)}^{\frac{1}{2}}} 
\end{equation}

From continuity of the geometry, we demand $V_{-} = V_{+}$, otherwise the geometry is not defined at the junction. From an integration around an infinitesimal region that contains $\phi = 0$ we obtain the jump of the $\xi$:
\begin{equation}\label{xid}
\Delta \xi = -\frac{V_{(\phi=0)}}{2\Omega_m} (\Lambda_{+\infty}-\Lambda_{-\infty})
\end{equation}
Inserting (\ref{xil}) into (\ref{xid}) gives the discontinuity of $\kappa$: 
\begin{equation}\label{kd}
\kappa_{+} - \kappa_{-} = -\frac{V_{(\phi=0)}^{\frac{3}{2}}}{2\Omega_m} (\Lambda_{+\infty}-\Lambda_{-\infty})
\end{equation}
Subtitling all the known term to the energy equation gives:
\begin{equation}
\begin{split}
E=\frac{1}{2}\dot{V}^2- \frac{D+d}{D+d-1}  V^2[\frac{2 \Omega_m}{V}(\xi+ \frac{1}{2})  \\+ \frac{\Lambda_{+\infty}-\Lambda_{-\infty}}{2} \textbf{Sign}(\phi) +\frac{\Lambda_{+\infty}+\Lambda_{-\infty}}{2}]
\end{split}
\end{equation}
Because of the jump of the potential and the field $\xi$, we can calculate the jump of $\dot{V}^2$ from the energy equation. The solution gives:
\begin{equation}
\frac{1}{2}\Delta\dot{V}^2=\frac{D+d}{D+d-1} (\frac{2 \Omega_m}{V_{(\phi=0)}}\Delta\xi +  \Lambda_{+\infty}-\Lambda_{-\infty}) V_{(\phi=0)}^2 = 0
\end{equation}
there is no jump in the volume and it's first derivative. From equation (\ref{sf}) which gives the dependence of the metric components we obtain that all derivatives of the scale factors are continuous. That leads to the conclusion that even when there a large discontinuous change in the potential, still the metric and it's derivative do not suffer from this discontinuities.
\subsection{Large times behavior and extra dimensional stabilization}
For obtaining the asymptotic limit of the solutions, lets take the case of pure vacuum energy $\Omega = \textrm{const}$ which is the case of late time expansion, we get an up side down harmonic oscillator, which for large volumes gives the solution:
\begin{equation}
V(t) = V_0 \exp(\chi t)
\end{equation}
where $\chi^2 = 2\frac{D+d}{D+d-1}\Omega$.
The integration form of the scale factors in Eq. (\ref{sf}) leads to:
\begin{subequations}
\begin{equation*}
R(t) \sim  e^{\frac{\chi t}{D+d}} \exp[-\sqrt{\frac{2Ed(D+d-1)}{D}} \frac{V_0^\frac{1}{D+d}}{\chi} e^{-\chi t}]
\end{equation*}
\begin{equation*}
r(t) \sim e^{\frac{\chi t}{D+d}} \exp[+\sqrt{\frac{2ED(D+d-1)}{d}} \frac{V_0^\frac{1}{D+d}}{\chi} e^{-\chi t}]
\end{equation*}
\end{subequations}
From the fact that the solution for integral gives $\exp(-\chi t)$ which decays for large times, we left with the limit:
\begin{subequations}
\begin{equation}
R(t) \rightarrow R_0 \exp(\frac{\chi}{D+d} t)
\end{equation}
\begin{equation}
r(t) \rightarrow r_0 \exp(\frac{\chi}{D+d} t)
\end{equation}
\end{subequations}
This represents a restoration of isotropy in the evolution of all dimensions in the universe. This has to be avoided, because the extra dimensions should be small also in the late universe. One way to archive this is to generate a potential for the extra dimensions which starts to act when the extra dimensions are very small, and then freeze the extra dimensions to very small size. This can be obtained, for example, by using the Casimir effect present in periodic extra dimensions  
\cite{Appelquist:1983vs}\cite{Appelquist:1988fh}\cite{Blau:1984wf}. The stopping of the extra dimensions can be used also as a particle production mechanism, that can result in the reheating of the universe by a field independent of the inflaton (our field $\phi$) which is the extra dimension size. The extra dimension size becomes therefore a curvaton field \cite{Enqvist:2001zp}\cite{Lyth:2001nq}\cite{Moroi:2001ct}.
\section{Discussion}
In this article we studied the basics of inflation compactification mechanism from the interplay of ordinary and higher dimensions. In the case of isotropic pressure, the solution can be obtained for the total volume and with no dependence with the individual scale factors of each dimension. Those can be calculated directly from the total volume dependence and the anisotropy constant $E$. 

For the dynamical space time theory, which produces a unification of dark energy, dark matter and a bounce of the volume, which naturally prevents the collapse of the universe and obtain a lower bound for the volume of the universe. Likewise the presence of a negative cosmological constant prevents the volume from becoming very big in the early universe. There is an effective potential that governs the evolution of the volume. In the case the effective potential is positive and has a minimum, a static solution for the total volume is obtained, and exponential compactification of the extra dimensions occurs. In that case the ordinary dimensions exponentially increase and the extra dimensions exponentially decrease. For small values of $E$ higher than the value obtained for the case the volume sits at the minimum, the total volume oscillates and the ordinary dimensions expand exponentially with an oscillatory modulation. 

The dynamical space time theory provides a natural way to exit from the inflation compactification epoch. The main reason for that is that theory allows two different potentials: $U(\phi)$ which drives directly the evolution of the scalar field $\phi$, and $V(\phi)$ which determines the value of the dark energy. It is therefore perfectly possible for the scalar field that drives the vacuum energy to smoothly climb into small positive values of vacuum energy, which is defined as the end of the inflation compactification. A semi analytic solution for step function potential is also derived. In this limit the matching of the solution at the value of the scalar field where the vacuum energy jumps, still respects the continuity of all components of the metric and also for it's time derivatives.

We have showed that for exponential growth of the total volume breaks the anisotropy, and all the scale factor start to expand in a similar fashion. Using the role of the inflation compactification mechanism, we have explained, is to push the extra dimensions to very low sizes and the ordinary dimension to very large sizes. However we cannot extend the model to all future time, since the vacuum energy at the end will restores the isotropy of the expansion of all dimensions. So we have to invoke a mechanism that locks the extra dimensions, when they reduce to very small sizes. This could be produced from the known Casimir effect that takes place in the compact extra dimension for example. The stopping of the extra dimensions can be used also as a particle production mechanism, that can result in the reheating of the universe by a field independent of the inflaton (our field $\phi$) which is the extra dimension size. The extra dimension size becomes therefore a curvaton field.

\section{identities}
The covariant conservation of the energy momentum tensor gives:
\begin{equation}
\dot{\rho} + (p+\rho) D \frac{\dot{R}}{R}+(p'+\rho)  \frac{\dot{r}}{r} = 0
\end{equation}
The Ricci tensor non vanishing values, under the metric (\ref{metric}):
\begin{equation}
R_{00} = -(D\frac{\ddot{R}}{R}+d\frac{\ddot{r}}{r})
\end{equation}
\begin{equation}
R_{DD} = \dot{H_D} + (DH_D+dH_d)H_D +(D-1)\frac{k_D}{R^2}
\end{equation}
\begin{equation}
R_{dd} = \dot{H_d} + (DH_D+dH_d)H_d +(d-1)\frac{k_d}{r^2}
\end{equation}
And the Ricci scalar:
\begin{equation}
\begin{split}
R = 2D \frac{\ddot{R}}{R} + 2d \frac{\ddot{r}}{r} + 2Dd H_DH_d \\+D(D-1)(H_D^2+\frac{k_D}{R^2}) +d(d-1)(H_d^2+\frac{k_d}{R^2})
\end{split}
\end{equation}
\acknowledgments
This article is supported by COST Action CA15117 "Cosmology and Astrophysics Network for Theoretical Advances and Training Action" (CANTATA) of the COST (European Cooperation in Science and Technology). In addition we thank the Foundational Questions Institute FQXi for support, in particular support for our conference BASIC2018, where this research was carried out.


\begin{thebibliography}{0}%
\makeatletter
\providecommand \@ifxundefined [1]{%
 \@ifx{#1\undefined}
}%
\providecommand \@ifnum [1]{%
 \ifnum #1\expandafter \@firstoftwo
 \else \expandafter \@secondoftwo
 \fi
}%
\providecommand \@ifx [1]{%
 \ifx #1\expandafter \@firstoftwo
 \else \expandafter \@secondoftwo
 \fi
}%
\providecommand \natexlab [1]{#1}%
\providecommand \enquote  [1]{``#1''}%
\providecommand \bibnamefont  [1]{#1}%
\providecommand \bibfnamefont [1]{#1}%
\providecommand \citenamefont [1]{#1}%
\providecommand \href@noop [0]{\@secondoftwo}%
\providecommand \href [0]{\begingroup \@sanitize@url \@href}%
\providecommand \@href[1]{\@@startlink{#1}\@@href}%
\providecommand \@@href[1]{\endgroup#1\@@endlink}%
\providecommand \@sanitize@url [0]{\catcode `\\12\catcode `\$12\catcode
  `\&12\catcode `\#12\catcode `\^12\catcode `\_12\catcode `\%12\relax}%
\providecommand \@@startlink[1]{}%
\providecommand \@@endlink[0]{}%
\providecommand \url  [0]{\begingroup\@sanitize@url \@url }%
\providecommand \@url [1]{\endgroup\@href {#1}{\urlprefix }}%
\providecommand \urlprefix  [0]{URL }%
\providecommand \Eprint [0]{\href }%
\providecommand \doibase [0]{http://dx.doi.org/}%
\providecommand \selectlanguage [0]{\@gobble}%
\providecommand \bibinfo  [0]{\@secondoftwo}%
\providecommand \bibfield  [0]{\@secondoftwo}%
\providecommand \translation [1]{[#1]}%
\providecommand \BibitemOpen [0]{}%
\providecommand \bibitemStop [0]{}%
\providecommand \bibitemNoStop [0]{.\EOS\space}%
\providecommand \EOS [0]{\spacefactor3000\relax}%
\providecommand \BibitemShut  [1]{\csname bibitem#1\endcsname}%
\let\auto@bib@innerbib\@empty
\end{thebibliography}%


\begin{thebibliography}{99}
\bibitem{Freund:1980xh} 
  P.~G.~O.~Freund and M.~A.~Rubin,
  Phys.\ Lett.\ B {\bf 97}, 233 (1980)
  [Phys.\ Lett.\  {\bf 97B}, 233 (1980)].
  doi:10.1016/0370-2693(80)90590-0
 \bibitem{Benisty:2018qed} 
  D.~Benisty and E.~I.~Guendelman,
  arXiv:1802.07981 [gr-qc].
   \bibitem{Gao:2009me} 
  C.~Gao, M.~Kunz, A.~R.~Liddle and D.~Parkinson,
  Phys.\ Rev.\ D {\bf 81}, 043520 (2010)
  doi:10.1103/PhysRevD.81.043520
  [arXiv:0912.0949 [astro-ph.CO]].
  \bibitem{Guendelman:1990kg} 
  E.~I.~Guendelman,
  `The inflation compactification mechanism,''
  LA-UR-90-2104.
  \bibitem{Guendelman:1993ty} 
  E.~I.~Guendelman and A.~B.~Kaganovich,
  Int.\ J.\ Mod.\ Phys.\ D {\bf 2}, 221 (1993).
  doi:10.1142/S0218271893000180
  \bibitem{Guendelman:2003tm} 
  E.~I.~Guendelman and A.~B.~Kaganovich,
  gr-qc/0302063.
    \bibitem{Ho:2010vv} 
  C.~M.~Ho and T.~W.~Kephart,
  Int.\ J.\ Mod.\ Phys.\ A {\bf 27}, 1250151 (2012)
  doi:10.1142/S0217751X12501515
  [arXiv:1002.4044 [hep-ph]].
  \bibitem{Szydlowski:1990ph} 
  M.~Szydlowski and M.~Biesiada,
  Phys.\ Rev.\ D {\bf 41}, 2487 (1990).
  doi:10.1103/PhysRevD.41.2487
\bibitem{Tosa:1984gr} 
  Y.~Tosa,
  Phys.\ Rev.\ D {\bf 30}, 2054 (1984)
  Erratum: [Phys.\ Rev.\ D {\bf 31}, 2697 (1985)].
  doi:10.1103/PhysRevD.30.2054, 10.1103/PhysRevD.31.2697

\bibitem{Kasner}
  Kasner, E. "Geometrical theorems on Einstein's cosmological equations." Am. J. Math. 43 (1921)
   \bibitem{Appelquist:1983vs} 
  T.~Appelquist and A.~Chodos,
  Phys.\ Rev.\ D {\bf 28}, 772 (1983).
  doi:10.1103/PhysRevD.28.772
   \bibitem{Appelquist:1988fh} 
  T.~Appelquist, A.~Chodos and P.~G.~O.~Freund,
  IN *APPELQUIST, T. (ED.) ET AL.: MODERN KALUZA-KLEIN THEORIES* 1-47
  \bibitem{Blau:1984wf} 
  S.~K.~Blau, E.~I.~Guendelman, A.~Taormina and L.~C.~R.~Wijewardhana,
  Phys.\ Lett.\  {\bf 144B}, 30 (1984).
  doi:10.1016/0370-2693(84)90170-9
  \bibitem{Enqvist:2001zp} 
  K.~Enqvist and M.~S.~Sloth,
  Nucl.\ Phys.\ B {\bf 626}, 395 (2002)
  doi:10.1016/S0550-3213(02)00043-3
  [hep-ph/0109214].
  \bibitem{Lyth:2001nq} 
  D.~H.~Lyth and D.~Wands,
  Phys.\ Lett.\ B {\bf 524}, 5 (2002)
  doi:10.1016/S0370-2693(01)01366-1
  [hep-ph/0110002].
  \bibitem{Moroi:2001ct} 
  T.~Moroi and T.~Takahashi,
  Phys.\ Lett.\ B {\bf 522}, 215 (2001)
  Erratum: [Phys.\ Lett.\ B {\bf 539}, 303 (2002)]
  doi:10.1016/S0370-2693(02)02070-1, 10.1016/S0370-2693(01)01295-3
  [hep-ph/0110096].
\end{thebibliography}
\end{document}